\documentclass[conference]{IEEEtran}
\IEEEoverridecommandlockouts
\usepackage{cite}
\usepackage{amsmath,amssymb,amsfonts}
\usepackage{algorithmic}
\usepackage{graphicx}
\usepackage{textcomp}
\usepackage{xcolor}
\usepackage{academicons}
\usepackage{orcidlink}
\usepackage{hyperref}

\usepackage{enumitem}
\def\BibTeX{{\rm B\kern-.05em{\sc i\kern-.025em b}\kern-.08em
    T\kern-.1667em\lower.7ex\hbox{E}\kern-.125emX}}
\usepackage{booktabs}

\newcommand{\orcid}[1]{\href{https://orcid.org/#1}{\textcolor{black}{\aiOrcid}}}

\begin{document}

\title{A First-Principles Based Risk Assessment Framework and the IEEE P3396 Standard}

\author{\IEEEauthorblockN{Richard J. Tong \orcidlink{0000-0003-2902-4156}}
\IEEEauthorblockA{\textit{Chair, IEEE Artificial Intelligence Standards Committee (AISC)} \\
\href{mailto:Richard.Tong@ieee.org}{Richard.Tong@ieee.org}
%\url{https://orcid.org/0000-0003-2902-4156}
}
~\\
\IEEEauthorblockN{Marina Cort\^es \orcidlink{0000-0003-0485-3767}}
\IEEEauthorblockA{\textit{Chair, IEEE P3395 Working Group;}\\ \textit{Institute of Astrophysics and Space Sciences, University of Lisbon, Portugal}\\ %\textit{Faculty of Sciences, Campo Grande, PT1749-016 Lisbon, Portugal}\\
\href{mailto:mcortes@ieee.org}{mcortes@ieee.org}
}
~\\
\IEEEauthorblockN{Jeanine A.\ DeFalco \orcidlink{0000-0002-8222-5005}}
\IEEEauthorblockA{\textit{Vice Chair, IEEE Artificial Intelligence Standards Committee}\\ \textit{University of New Haven}\\
\href{mailto:drj9defalco@gmail.com}{drj9defalco@gmail.com}
}
~\\
\IEEEauthorblockN{Mark Underwood %\orcidlink{0000-0000-0000-0000}
}
\IEEEauthorblockA{\textit{Chair, IEEE P3396 Working Group}\\
\href{mailto:knowlengr@gmail.com}{knowlengr@gmail.com}}
~\\
\IEEEauthorblockN{Janusz Zalewski \orcidlink{0000-0002-2823-0153}}
\IEEEauthorblockA{\textit{Secretary, IEEE P3395 Working Group;}\\ \textit{Florida Gulf Coast University, United States;}\\
\textit{State Academy of Applied Sciences, Ciechanow, Poland}\\
\href{mailto:ikswelaz@gmail.com}{ikswelaz@gmail.com}
}}
%\url{https://orcid.org/0000-0003-0485-3767}

\maketitle

\begin{abstract}
Generative Artificial Intelligence (AI) is enabling unprecedented automation in content creation and decision support, but it also raises novel risks. This paper presents a first-principles risk assessment framework underlying the IEEE P3396 Recommended Practice for AI Risk, Safety, Trustworthiness, and Responsibility. We distinguish between \emph{process risks} (risks arising from how AI systems are built or operated) and \emph{outcome risks} (risks manifest in the AI system's outputs and their real-world effects), arguing that generative AI governance should prioritize outcome risks. Central to our approach is an \textbf{information-centric ontology} that classifies AI-generated outputs into four fundamental categories: (1) \emph{Perception-level information}, (2) \emph{Knowledge-level information}, (3) \emph{Decision/Action plan information}, and (4) \emph{Control tokens} (access or resource directives). This classification allows systematic identification of harms and more precise attribution of responsibility to stakeholders (developers, deployers, users, regulators) based on the nature of the information produced. We illustrate how each information type entails distinct outcome risks (e.g. deception, misinformation, unsafe recommendations, security breaches) and requires tailored risk metrics and mitigations. By grounding the framework in the essence of information, human agency, and cognition, we align risk evaluation with how AI outputs influence human understanding and action. The result is a principled approach to AI risk that supports clear accountability and targeted safeguards, in contrast to broad application-based risk categorizations. We include example tables mapping information types to risks and responsibilities. This work aims to inform the IEEE P3396 Recommended Practice and broader AI governance with a rigorous, first-principles foundation for assessing generative AI risks while enabling responsible innovation.
\end{abstract}


\begin{table*}[ht]\centering
\caption{Categories of AI-Generated Information with Examples}
\begin{tabular}{p{3.3cm}p{9.5cm}}
\toprule
\textbf{Information Type} & \textbf{Example Output and Description} \\
\midrule
\emph{Perception-level} 
& AI-generated content that simulates perceptual experience. \textit{Example:} a deepfake video or synthetic image purporting to show a real person or event (manipulating what viewers perceive). \\
\addlinespace[0.6em]
\emph{Knowledge-level} 
& AI-generated informative content. \textit{Example:} an answer to a factual question or a summary of a topic produced by an AI (intended to convey knowledge or truth about the world). \\
\addlinespace[0.6em]
\emph{Decision/Action plan} 
& AI output that recommends actions or makes decisions. \textit{Example:} a generative model’s advice to a user (“Based on your symptoms, you should take medication X”), or an AI system’s autonomous decision plan (like a strategy for a robot or a loan approval decision). \\
\addlinespace[0.6em]
\emph{Control tokens} 
& AI-generated tokens that directly control access or operations. \textit{Example:} a password, API key, or command sequence produced by an AI (which, if used, can grant access to a system or execute an action, e.g. code that controls a device). \\
\bottomrule
\end{tabular}
\label{tab:infotypes}
\end{table*}

\section{Introduction}
Generative AI systems (such as large language models and generative image models) have sparked both excitement and concern \cite{NatAcad,Suleyman,OECD,AIsafety,P3395,P3395-III}. As these systems proliferate in society, a key challenge is how to evaluate and manage their risks in a principled way. We advocate a \textbf{first-principles approach} to AI risk assessment: rather than starting from technology categories or use-case labels, we begin by defining risk in terms of fundamental causes and effects. In risk management theory, risk is often defined as the probability of an adverse event multiplied by its impact severity \cite{NIST2023}. Applying this to AI, we consider what adverse events can result from an AI system and how severe those outcomes could be. Critically, we distinguish between two kinds of risk in AI systems:
\begin{itemize}\itemsep 0pt
    \item \emph{Process risk}: the risk arising from \textit{how} the AI is developed or operated. This includes issues like biased training data, inadequate testing, lack of transparency, or poor model oversight during development and deployment.
    \item \emph{Outcome risk}: the risk arising from \textit{what} the AI does or produces in the real world. This refers to the direct harms or adverse impacts caused by the AI’s outputs or actions.
\end{itemize}
Traditional AI governance has often emphasized process (e.g. mandating documentation, algorithmic transparency, or bias audits during development). While managing process risk is important, a first-principles analysis suggests that for generative AI, \textbf{outcome risks should be the primary focus}. Ultimately, what matters is the effect an AI system has: a perfectly transparent model can still generate harmful content, and conversely a system built with imperfect processes might pose little danger if its outputs are benign. We aim to evaluate AI systems based on \emph{actual consequences} rather than abstract fears or broad labels.

This perspective aligns with an emerging consensus to regulate AI according to its \textit{use and impact} rather than simply its category or technology. For example, the European Union’s proposed AI Act defines risk tiers by application areas (e.g. “AI in education” or “biometric identification” as high-risk sectors). Critics have noted that such broad categorization can be both over-inclusive and under-inclusive, failing to account for the specific ways an AI causes harm in context \cite{EUAIAct,Veale2021}. Instead, our approach looks at the AI system’s outputs and asks: \emph{how could this output cause harm, and who can prevent it?} The IEEE P3396 working group has adopted this philosophy, developing a principles-based framework that examines the role of AI in information generation, decision-making, human agency, and responsibility throughout the AI lifecycle.

In this paper, we articulate the \textbf{information-centric risk assessment framework} underpinning IEEE P3396. We introduce an ontology that classifies AI-generated information into four fundamental types, corresponding to different levels of semantic content and effect on human decision processes. By analyzing generative AI through the lens of the information it produces, we can tie potential harms directly to those outputs and assign responsibility to the appropriate stakeholders by first principles. This allows more precise risk attribution than approaches that classify AI systems only by domain or general functionality. We also delineate how outcome risks differ from process risks for each information category, and how governance measures can address each. Table~\ref{tab:infotypes} summarizes the four information types with examples. Subsequent sections will examine the outcome vs.\ process risk considerations (Table~\ref{tab:risks}) and responsibility allocation (Table~\ref{tab:responsibility}) for each category. We then discuss the philosophical and theoretical grounding for this framework in concepts of information, cognition, and agency, before concluding with implications for AI standards and policy.

\begin{table*}[ht]\centering
\caption{Outcome vs. Process Risks for Different AI Output Types}
\begin{tabular}{p{3.2cm} p{5.8cm} p{5.8cm}}
\toprule
\textbf{Output Type} & \textbf{Primary Outcome Risks (Direct Harms)} & \textbf{Key Process Risk Factors} \\
\midrule
\emph{Perception-level} 
& Deception, manipulation of beliefs, reputational harm, psychological distress. (e.g. deepfake videos causing false beliefs or harassment) 
& Inadequate content filters allowing realistic fake media; biased training data leading to stereotyped or defamatory outputs; lack of watermarks or provenance tracking. \\
\addlinespace[0.8em]
\emph{Knowledge-level} 
& Misinformation, erosion of trust in information, bad decisions based on false data. (e.g. hallucinated facts in an AI-generated report lead a user to a wrong conclusion) 
& Poor training data quality or verification (model “hallucinations” of false facts); lack of updates or fact-checking modules; model bias causing systematic factual errors or one-sided information. \\
\addlinespace[0.8em]
\emph{Decision/Action plan} 
& Unsafe or unfair outcomes from following AI advice or decisions: physical harm, financial loss, discrimination, violation of rights. (e.g. a flawed medical recommendation causes harm; an automated decision system unjustly denies services) 
& Inadequate constraint of AI recommendations (no safety limits or ethical guidelines in training); insufficient testing in the deployment domain; bias in training data leading to unfair decisions; lack of human oversight for critical decisions. \\
\addlinespace[0.8em]
\emph{Control tokens} 
& Security breaches, privacy violations, unauthorized actions. (e.g. AI reveals a private password or generates malicious code that can be executed) 
& Model memorization of sensitive data from training (e.g. secrets included in training set and reproduced); absence of filters to detect and remove secret keys or dangerous code; lack of access control in deployment (AI directly interfacing with critical systems without safeguards). \\
\bottomrule
\end{tabular}
\label{tab:risks}
\end{table*}

\section{An Information-Centric Framework for Generative AI Risk}
\label{sec:framework}
\subsection{Categories of AI-Generated Information}
Generative AI systems can produce a wide range of outputs. We propose that nearly all AI-generated \emph{informational artifacts} can be classified into four broad categories, based on the nature of information they represent and how that information is used by humans or other systems. These categories, listed in Table~\ref{tab:infotypes}, are: 
\textbf{(1) Perception-level information}, \textbf{(2) Knowledge-level information}, \textbf{(3) Decision/Action plan information}, and \textbf{(4) Control tokens}. 

Each category corresponds to a distinct layer of semantic content and potential impact:
\begin{enumerate}\itemsep 0pt
    \item \textbf{Perception-level information} refers to AI-generated content intended to be consumed by human senses or treated as direct observations. This includes images, video, audio, or descriptive text that portrays events or scenarios. Such outputs function at the level of \emph{perception}—they can alter or manipulate what people see and hear, and thus what they believe to have happened. For example, an AI-generated photograph or deepfake video is perception-level information: it provides a visual depiction that viewers might interpret as real. 
    \item \textbf{Knowledge-level information} refers to content that conveys facts, data, or answers to questions—informational outputs meant to inform or describe reality. This is the level of \emph{propositional knowledge}. Examples include a paragraph generated by a language model explaining a scientific concept, an AI-written news article, or answers in a Q\&A system. The goal of such output is to impart knowledge (true or false) to the user. It is akin to an encyclopedia entry or an expert’s report provided by the AI.
    \item \textbf{Decision/Action plan information} is content that recommends or leads to actions. This category covers AI outputs that either \emph{advise a decision} (e.g. a recommendation, prediction or evaluation that a human might act on) or \emph{formulate an action plan} (e.g. instructions, strategies, or decisions made by the AI). It spans both non-binding advice (“You should invest in X stock”) and more explicit directives or plans (“Here is a step-by-step plan to treat this illness” or an AI system’s decision to approve a loan). In essence, this is information explicitly geared towards influencing choices and behaviors.
    \item \textbf{Control tokens (access/resource directives)} are a special category of AI output consisting of unique codes or signals that directly enact control over systems or resources. These include outputs like passwords, cryptographic keys, authentication tokens, or command codes. They may be generated by AI intentionally (e.g. an AI developer tool generating an API key or code snippet) or even unintentionally (an AI model regurgitating a sequence from its training data that happens to be a password or personal identifier \cite{Carlini2021}). Such outputs have \emph{operational significance} beyond human interpretation: they can grant access, identify an individual, or trigger an automated process.
\end{enumerate}

These categories capture the \textit{level of information} the AI output embodies, drawing a distinction between content aimed at human perception, content conveying factual knowledge, content guiding decisions/actions, and content that serves as an operative token. The categorization is not strictly exclusive (a single AI output could contain multiple types of information), but it provides a useful analytical lens. By identifying what type(s) of information an AI system generates, we can more directly infer \textbf{what kind of harm it might cause}. For instance, if an AI produces perception-level content (like images), we immediately consider risks like visual deception or defamation; if it produces decision-level content, we consider the consequences of following incorrect or biased advice.

Notably, this information-centric taxonomy corresponds in part to the classic cognitive pipeline: perception leads to knowledge, which informs decisions, which lead to actions. It is rooted in how information flows through cognitive agents \cite{Newell1982,Russell2021}. In the context of AI, an autonomous system or a human-AI team can be analyzed by the information the AI contributes at each stage (perceptual input, knowledge inference, decision suggestion, or direct control signals). By classifying outputs along these lines, we ground our risk analysis in the fundamental ways AI technology interfaces with human cognition and agency.

\subsection{Outcome Risks Across Information Categories}
Each category of AI-generated information comes with a characteristic set of \emph{outcome risks} — the direct harms or adverse outcomes that could result if the information is flawed, misused, or maliciously employed. Table~\ref{tab:risks} outlines the primary outcome risks associated with each information type, along with the relevant \emph{process risk factors} that might contribute to those outcomes.

Focusing on outcome risks means concentrating on the actual harm scenarios for each output type. For perception-level information, a prominent risk is \emph{deception}: AI-generated media can fool people into believing something occurred which did not. A deepfake video, for example, might depict a public figure saying something they never said, potentially swaying public opinion or damaging reputations. Such output can facilitate fraud or propaganda, leading to societal harms (as evidenced by the prevalence of deepfake pornography and misinformation \cite{Chesney2019}). The process factors that heighten this risk include whether the AI developer implemented safeguards like deepfake detection or watermarks, and whether the deployer has policies to prevent malicious use (e.g. disallowing use of a face-swapping tool without consent). If those process measures are lacking, the likelihood of deceptive outcomes increases.

\begin{table*}[ht!]\centering
\caption{Primary Risk Mitigation Focus and Responsible Parties by Output Type}
\begin{tabular}{p{3.2cm} p{4.8cm} p{6cm}}
\toprule
\textbf{Output Type} & \textbf{Key Risk Metric/Focus} & \textbf{Primary Responsible Stakeholders} \\
\midrule
\emph{Perception-level} 
& Authenticity and consent. (Preventing deceptive or harmful manipulations of media; ensuring subjects’ rights are respected in generated content.) 
& \textbf{Developer}: implement content filters, watermarking of synthetic media, detection of deepfakes. \newline
\textbf{Deployer}: enforce usage policies (no non-consensual image generation, user identity verification for sensitive uses). \newline
\textbf{User}: refrain from malicious use or unwarranted trust in unverified media. \\
\addlinespace[1.0em]
\emph{Knowledge-level} 
& Accuracy and veracity. (Maximizing truthfulness of outputs; minimizing false or misleading information.) 
& \textbf{Developer}: improve model training and prompting to reduce hallucinations, bias, and errors; possibly incorporate citation and fact-checking in the AI. \newline
\textbf{Deployer}: scope the application to appropriate domains, provide disclosures or warnings (e.g. “AI-generated content may be incorrect”), and facilitate user verification (like linking to sources). \newline
\textbf{User}: verify critical information with trusted sources, especially in high-stakes scenarios. \\
\addlinespace[1.0em]
\emph{Decision/Action plan} 
& Safety and fairness of decisions. (Avoiding harmful or unjust advice and decisions; ensuring alignment with ethical and legal norms in actions taken.) 
& \textbf{Developer}: build in guardrails against clearly dangerous recommendations (e.g. refuse instructions for self-harm or violence) and against discriminatory decision-making (ensure training data diversity and bias mitigation). \newline
\textbf{Deployer}: rigorously test AI decisions or advice in the specific context; if used in high-stakes domains (medicine, finance, hiring), include human oversight or review; comply with sector-specific regulations (e.g. require informed consent for AI advice, audit automated decisions for fairness). \newline
\textbf{User}: maintain human agency by treating AI advice as input rather than gospel—especially for important decisions, cross-check with human experts; use AI tools as decision support, not autonomous decision-makers, unless properly vetted. \newline
\textbf{Regulator}: set guidelines for AI in sensitive uses (e.g. requiring transparency that a decision was AI-assisted and avenues for recourse or appeal). \\
\addlinespace[1.0em]
\emph{Control tokens} 
& Security and privacy. (Preventing unauthorized access or malicious actions due to AI output; protecting sensitive information.) 
& \textbf{Developer}: implement methods to prevent the AI from leaking memorized secrets or generating dangerous code (e.g. penalize outputs that match known keys or exploit patterns); include robust authentication steps before any AI-generated command can directly affect critical systems. \newline
\textbf{Deployer}: if the AI is connected to system controls or data, put strict limits on its autonomy (sandbox execution of AI-written code, review AI actions before deployment in real-world systems); filter outputs for any patterns that look like keys or personal data; ensure compliance with privacy regulations. \newline
\textbf{User}: avoid using AI to intentionally retrieve sensitive data one is not authorized to access (such misuse is culpable); treat any AI-provided code or token with caution and verify its safety before use. \newline
\textbf{Regulator}: may mandate audits for systems that can impact cybersecurity or critical infrastructure via AI outputs, similar to safety certifications. \\
\bottomrule
\end{tabular}
\label{tab:responsibility}
\end{table*}

For knowledge-level outputs, the core outcome risk is \emph{false or misleading information}. Users may take AI-generated text or answers as true, so any hallucination or error can propagate misinformation. The real-world impact ranges from trivial (minor factual errors) to severe (e.g. incorrect medical or legal information leading to bad decisions). A notable example is an AI language model fabricating legal case citations that a lawyer then submitted in a court brief, resulting in sanctions when the falsehood was exposed. The accuracy and reliability of knowledge outputs are paramount. Process factors that contribute to risk include the thoroughness of training data curation and model validation for truthfulness, and whether mechanisms like fact-checking or citation are built into the system. A model trained on unverified data without oversight is more likely to produce confident-sounding falsehoods.

Decision and action plan outputs have outcome risks directly tied to the consequences of acting on AI advice or delegating decisions to AI. If a generative model gives poor financial advice, a user could lose money; if an AI decision system used in hiring is biased, qualified candidates could face unfair rejection, implicating their rights. These \emph{outcome risks} often correspond to what regulators call high-risk AI applications (those affecting health, safety, or fundamental rights) \cite{EUAIAct}. Generative AI expands this concern because even a general-purpose model not specifically built for high-stakes decisions can \emph{influence} decisions (e.g. a chatbot giving unvetted medical or legal guidance). Thus the outcome risk is that AI advice or decisions are taken improperly. Key process safeguards here include aligning the AI during development to avoid giving dangerous or unethical instructions (many chatbots refuse certain advice), careful domain-specific testing by deployers, and integrating human oversight for critical uses. Failure in these processes (for instance, deploying an AI system in a medical context without validation or user training) raises the probability of a harmful decision outcome.

Control token outputs present perhaps the most unique set of risks. The outcome harms revolve around breaches of security or misuse of systems: if an AI inadvertently leaks a password or key, unauthorized parties might gain access to confidential data; if an AI generates exploit code or hazardous instructions and these are executed, it could cause direct damage. One documented concern is that large generative models have memorized and output sensitive sequences from their training data (including API keys and personal identifiers) \cite{Carlini2021}. The risk is fundamentally about confidentiality and integrity. Process measures to mitigate this include techniques to reduce memorization of training data or to detect and redact any sensitive patterns before output. Additionally, if an AI is allowed to directly control systems (e.g. code generation that is immediately executed in automation), strict sandboxing and review are process requirements to prevent unintended harmful actions. Skimping on these precautions can turn a normally low-risk AI (e.g. a coding assistant) into a source of serious outcome risk (if it ends up producing a command that wipes data or a token that grants admin privileges, and those are acted on blindly).

By mapping outcome risks in this way, we adhere to a first-principle question: \emph{what can go wrong when someone uses this AI’s output?} For each category of information, the answer is different, and that guides us to different safety strategies. Importantly, this approach avoids treating “AI” as monolithically risky or not; it recognizes, for example, that an image generator and a decision-making AI pose different kinds of dangers and thus need different guardrails. It also highlights the connections between process and outcome: a flaw in development (process risk) is concerning primarily because of the outcome it might lead to. By understanding the chain from process to outcome for each information type, we can prioritize interventions that break that chain and prevent harm.

\section{Risk Attribution and Stakeholder Responsibility}
A robust risk framework must also address \textbf{who is responsible} for managing and mitigating the identified risks. AI systems often involve multiple parties—from those who build the underlying model, to those who deploy it in a specific context, to the end users and those impacted by its outputs. Clear attribution of responsibility is essential to avoid gaps in accountability (sometimes referred to as the “responsibility gap” problem for autonomous systems \cite{Matthias2004}) and to ensure each stakeholder takes appropriate actions to reduce risk.

We identify four key stakeholder roles in the AI ecosystem \cite{Whittlestone2021}: 
\emph{developers} (who create the AI model or algorithm), 
\emph{deployers} (who integrate the AI into a product or application and offer it to users), 
\emph{users} (who interact with the AI system and apply its outputs), and 
\emph{regulators/oversight bodies} (who establish rules or standards and may intervene when necessary). 
These roles carry different responsibilities depending on the type of information output. Table~\ref{tab:responsibility} summarizes the primary risk mitigation focus and responsibility for each category of AI output.

The above allocation makes clear that different kinds of AI outputs shift the focus of risk management. For perception-level content (e.g. deepfake images), much of the onus is on \textbf{developers} and \textbf{deployers} to ensure tools are not easily misused to produce illicit material. For example, developers can integrate watermarking to mark images as AI-generated, and deployers can require user verification to prevent anonymous abuse of a generative image platform. Users, in turn, have a responsibility to not knowingly spread AI-generated falsehoods or violate others’ privacy with AI-created media. If a harmful deepfake does emerge, responsibility can be parsed: a user who maliciously created or distributed it is directly culpable, but if the platform (deployer) lacked reasonable safeguards, they share blame, and if the developer knowingly released a model prone to such misuse (e.g. without any content moderation capabilities), they too bear responsibility. This disentanglement helps avoid a “responsibility gap” \cite{Matthias2004} where each party deflects blame; instead, each has a clear role in prevention.

For knowledge-level outputs, \textbf{accuracy} is the key metric, and both developers and deployers must collaborate to achieve it. The developer of a large language model should strive to minimize hallucinations (perhaps using reinforcement learning from human feedback to favor correct answers). The deployer who fine-tunes or applies the model in a domain like medicine or law must validate its answers against known references and perhaps constrain it (e.g. an AI medical assistant might be limited to providing information from an approved medical database). Regulators might not be heavily involved in general knowledge applications, but could step in if there is systematic deception (for instance, consumer protection agencies may act if a company’s AI consistently provides dangerously false information in a product). Users are expected to exercise critical thinking—just as one would not trust an anonymous blog without verification, one should double-check important information from an AI. However, recognizing that users may overestimate AI’s accuracy, the burden is primarily on providers to warn and design the system responsibly.

Decision and action-oriented outputs require the greatest oversight due to their potential to cause immediate harm. Here, the \textbf{deployer} often has the primary accountability for outcomes in context. If a hospital deploys an AI diagnostic tool, it must ensure the tool has passed rigorous clinical validation and that doctors understand its limitations. The \textbf{developer} must furnish the AI with necessary safety features (not providing certain advice, flagging uncertainty, etc.) and transparency to facilitate this oversight. \textbf{Regulatory bodies} frequently impose requirements in such scenarios, such as documentation of algorithmic decision criteria, audit trails, and the right for an affected individual to contest an automated decision. The EU AI Act, for example, considers many decision-support AIs in fields like hiring or credit to be “high-risk” and subject to specific obligations \cite{EUAIAct}. Our framework refines this by indicating that it’s not just the field (education, employment, etc.) but the type of information output (a recommendation or decision) that triggers the need for such controls. The end \textbf{users} (or subjects) of AI decisions should be empowered with information and recourse—maintaining human agency means an AI’s decision should usually be advisory or subject to human confirmation when stakes are high. In cases where users deliberately use an AI for illicit decision support (e.g. to plan a crime), those users carry full moral and legal responsibility, akin to using any tool for wrongdoing.

Finally, for control token outputs, \textbf{security} is paramount. If an AI system can output actual keys or commands, the \textbf{deployers} and \textbf{developers} must treat it almost like developing a cybersecurity product. Any AI integrated into system operations should undergo threat modeling to prevent it from becoming an attack vector (e.g. prompting it in a way that makes it leak admin passwords). Strong safeguards (both technical and procedural) are needed, and regulators may classify certain AI capabilities (like code generation that can control critical systems or chemical formula generation that could be dangerous) as sensitive technology subject to export controls or special handling \cite{Brundage2018}. In less extreme cases, simply ensuring that AI-generated tokens are treated with the same caution as human-generated ones is key; for instance, if an AI outputs software code, that code should be reviewed and tested as thoroughly as if a human wrote it before being deployed. Users must avoid complacency—just because an AI produced a piece of code or a configuration does not guarantee its safety. The recent emphasis on “human in the loop” for AI in complex systems is essentially about keeping a responsible human agent accountable for any action taken by the AI, particularly when those actions can have far-reaching effects.

In summary, our framework’s classification not only pinpoints \emph{what} could go wrong (outcome risks), but also clarifies \emph{who} should act to prevent it. By attributing responsibilities aligned with the type of information output, we ensure that each stakeholder has actionable duties: developers focus on intrinsic safety of the AI outputs, deployers focus on safe use in context, users exercise judgment and compliance, and regulators set boundaries where needed to protect the public. This structured allocation is vital to operationalize AI risk management in practice and is a core principle in the development of IEEE P3396. It aims to foster \textbf{trustworthy AI} by designating clear lines of accountability, which in turn incentivizes each stakeholder to uphold their part in risk mitigation.

\section{Theoretical Foundations: Information, Cognition, and Agency}
Underpinning our risk framework are foundational principles from information theory, cognitive science, and philosophy of technology. We briefly highlight how these ideas inform the categorization and risk attribution approach:

\paragraph{The primacy of information:} Claude Shannon’s seminal work established information as a measurable entity, separate from its meaning, that underlies communication systems \cite{Shannon1948}. Building on this, philosophers like Dretske and Floridi have examined the \emph{semantic} nature of information—how signals convey meaning and knowledge \cite{Dretske1981, Floridi2011}. Our framework implicitly leverages both the quantitative and qualitative aspects of information. By focusing on AI outputs as information artifacts, we treat the AI system largely as an information source whose outputs can be analyzed for truth, relevance, and potential impact. The categories (perception-level, knowledge-level, etc.) reflect differing \emph{semantic content} and roles of information. This resonates with the Data–Information–Knowledge hierarchy in epistemology, where raw perceptual data is distinguished from processed information and actionable knowledge \cite{Ackoff1989}. In effect, we are categorizing AI outputs along an epistemic spectrum from mere appearance (perceptual data) to instructive propositions (knowledge and decisions) to instrumental commands (control tokens). This explicit focus on information types is a first-principles stance that the essence of what AI produces is \emph{information}, and thus any risk analysis must start there, before considering the physical or social container around it.

\paragraph{Cognition and levels of analysis:} In cognitive science and AI, it is common to describe systems in terms of hierarchical levels (perception, cognition, action). Newell’s concept of the “knowledge level” in AI, for instance, suggests analyzing intelligent agents by the knowledge they have and use, abstracting away from implementation details \cite{Newell1982}. Our knowledge-level information category aligns with treating AI outputs at that abstract level of content, distinct from how the AI internally generated it. Meanwhile, perception-level and action-level (decision) categories correspond to input and output stages of a cognitive agent. Marr’s levels of analysis (computational, algorithmic, implementational) also emphasize that one can understand a system by the problem it solves (e.g. providing truthful answers) separately from the mechanism (neural network weights, etc.). By categorizing outputs, we operate mostly at Marr’s “computational” and “algorithmic” levels, focusing on what information task the AI output fulfills (describe an image, answer a question, suggest an action) rather than the hardware or model specifics. This abstraction ensures our risk framework is general: it could apply to a GPT-4, or a future more advanced AI, or even a non-AI information system, as long as the output role is the same.

\paragraph{Agency and responsibility:} A central concern in AI ethics is the notion of agency—who (or what) is the agent causing an outcome, and thus who can be held accountable. Philosophers like Andreas Matthias have pointed out the “responsibility gap” that can occur if we consider an autonomous system as having acted, yet no human is held responsible for its actions \cite{Matthias2004}. Our framework addresses this by explicitly maintaining that AI systems, especially generative ones, are \emph{tools that extend human agency} but do not eliminate it. By dissecting the chain of information production and usage, we ensure a human agent is identified at each link: the developer as the agent behind the AI’s design, the deployer as the agent deciding its use and context, the user as the agent deciding to act (or not) on its output. We have deliberately avoided framing the AI itself as an agent with responsibilities, even though AI can appear to “decide” or “act.” Instead, in our taxonomy, what the AI provides (e.g. a decision suggestion) is \emph{information input to a human or organizational decision-making process}. This viewpoint is philosophically aligned with researchers like Bryson who argue that AI should be treated as tools (“artefacts”) for which humans remain the moral agents \cite{Bryson2017}. By anchoring responsibility in human roles, we uphold the principle that advancing AI should not dilute human accountability. The IEEE P3396 standard embraces this, promoting human-centric governance where technologies serve human goals and humans remain answerable for outcomes.

\paragraph{Ethics of outcomes vs processes:} There is a broader philosophical debate in ethics and law about outcome-based versus deontological (duty/process-based) perspectives. In regulating AI, one sees echoes of this: should we judge an AI system by its internal adherence to certain principles (transparency, fairness in design) or by the consequences it produces? Our stance is largely consequentialist in the context of risk: we focus on outcomes (harms or benefits) to determine risk levels. This does not mean process is irrelevant — indeed, we identify process risk factors that can lead to bad outcomes. But the moral weight in our framework is given to preventing harm. This is in line with utilitarian approaches to technology ethics that prioritize minimizing suffering and maximizing benefits \cite{Bostrom2014}. It also reflects pragmatic regulatory thinking: policy interest in AI is driven by what AI does to people (discrimination, misinformation, accidents) rather than the elegance or opacity of the algorithms per se. That said, our framework’s information-centric approach bridges to duty-based ethics too: for example, ensuring an AI provides truthful information can be seen as fulfilling a duty of honesty, and ensuring an AI does not produce non-consensual deepfakes upholds duties of respect and privacy. In this way, articulating explicit information-based categories helps map general ethical principles (truthfulness, non-maleficence, justice, autonomy) onto concrete design and deployment requirements for each output type.

\paragraph{Informing standards and policy:} Lastly, we note that the structure of our framework is meant to be \emph{action-guiding} for standards like IEEE P3396. The philosophy of technology standards often emphasizes consensus on best practices and the translation of high-level values into implementable requirements \cite{Winfield2019}. By organizing risk by information outputs, we create a clear linkage from fundamental values (e.g. truth, safety, fairness, privacy) to technical and procedural measures. Each category of output encapsulates certain values at risk: perception outputs relate to truth (avoiding deception) and dignity (avoiding harmful portrayal), knowledge outputs to truth and knowledge integrity, decision outputs to fairness and safety, and control tokens to security and privacy. This mapping provides a first-principles justification for why particular standards or controls are needed: for instance, why watermarking (a control for perception-level output integrity) is ethically necessary, or why auditing an AI’s decisions for bias (a control for decision-level fairness) is required. Rather than enumerating endless “AI use cases” and adding piecemeal rules, our framework offers a principled scaffolding that can cover future AI systems by focusing on their informational essence. We see this as philosophically robust and practically flexible, which is crucial in a rapidly evolving field.

\section{Conclusion}
We have presented a first-principle based risk assessment framework for generative AI, structured around the fundamental types of information these systems produce. By categorizing AI outputs into perception-level, knowledge-level, decision/action, and control token information, we achieve a more precise understanding of potential harms and responsibilities than broad-brush approaches. This information-centric view aligns closely with how AI affects human cognition and action: each category pinpoints a stage where AI-generated information can influence beliefs or behavior, and thereby highlights specific risk mitigation needs.

Our framework distinguishes outcome risks from process risks, emphasizing that while good development practices are essential, it is ultimately the outcomes (the AI’s effects in the world) that justify regulatory and ethical concern. Focusing on outcomes leads us to prioritize measures like preventing deceptive content, ensuring informational accuracy, safeguarding decision integrity, and maintaining security — all tailored to the kind of content the AI generates. In doing so, we avoid one-size-fits-all labels of “high-risk AI” and instead evaluate risk with granularity and context.

Crucially, this approach also clarifies the allocation of responsibility. We assert that AI does not obviate human agency: developers, deployers, and users each have distinct duties to ensure AI is used safely and ethically. By linking those duties to the information types (for example, developer’s duty to curb hallucinations in knowledge outputs, deployer’s duty to oversee AI decisions in sensitive domains, user’s duty not to misuse control information), we close potential responsibility gaps. This clarity is not only philosophically satisfying, but it provides practical guidance for governance: it tells organizations and regulators \emph{who should do what} to manage AI risk.

The work reported in this paper is intended to inform the ongoing development of the IEEE P3396 Recommended Practice. In a standards context, our framework can translate into checklists or requirements for AI system assessments. For instance, an assessment procedure could ask: “Does the AI output perception-level information? If so, implement content provenance and misinformation countermeasures. Does it output decision-level information? If so, implement oversight and fairness audits,” and so on for each category. By following a first-principles structure, such a standard can cover known and yet-to-be-invented AI applications, since it rests on timeless aspects of information and agency.

Future work will involve validating and refining this framework through case studies and stakeholder feedback. Scenarios in education, healthcare, finance, etc., can be analyzed with our approach to ensure it captures all relevant risks and is easily applicable. Additionally, as AI technology evolves (e.g. more advanced multimodal systems, agents that can act autonomously), we will examine if new information categories emerge or if the existing ones suffice. So far, the categories proposed have shown versatility, but continual re-evaluation is a principle of first-principles thinking itself — to adapt our understanding as reality evolves.

In conclusion, the information-centric risk assessment framework offers a promising path to AI governance that is both principled and practical. It speaks to foundational truths about information and human responsibility, while providing concrete tools for risk analysis. We believe this approach will enhance the ability of organizations and regulators to ensure AI systems are safe, trustworthy, and aligned with human values. By adopting such first-principle frameworks, we can better reap the benefits of generative AI while guarding against its pitfalls, laying a stable foundation for standards like IEEE P3396 and beyond.

\section*{IEEE disclaimer}
This IEEE AISC (``IEEE-AISC'') publication (``Work'') is {\it not} a consensus standard document. Specifically this document is NOT AN IEEE STANDARD.
This article solely represents the views of the set of authors within the IEEE AISC P3396 Working Group, and does not necessarily represent a position of either IEEE or the IEEE Standards Association.
  Information contained in this ``Work'' has been created by, or obtained from, sources believed to be reliable, and reviewed by members of the activity that produced this ``Work''.
 Although the WG members who have created this Work believe that the information and guidance given in this ``Work'' serve as an enhancement to users, all persons must rely upon their own skill and judgment when making use of it.

\end{document}